\documentclass[reprint,aip,pop]{revtex4-1}

\usepackage{graphicx}
\usepackage{amsmath}
\usepackage{amssymb}
\usepackage{color}

\begin{document}


\title{Backward Raman amplification in the Langmuir wavebreaking regime} 
\author{Z.~Toroker} 
\affiliation{Department of Electrical Engineering, Technion Israel Institute of Technology, Haifa 32000, Israel}
\author{V.~M.~Malkin}
\author{N.~J.~Fisch} 
\affiliation{Department of Astrophysical Sciences, Princeton University, Princeton, NJ USA 08540}



\date{\today}

\begin{abstract}


In plasma-based backward Raman amplifiers, the output pulse intensity increases with the input pump pulse intensity,
 as long as the Langmuir wave mediating energy transfer from the pump to the seed pulse remains intact. 
However,  at high pump intensity,  the Langmuir wave breaks, at which point  the amplification efficiency  may no longer increase with the pump intensity. 
Numerical simulations presented here, employing a 1D Vlasov-Maxwell code,  show that, although the amplification efficiency remains high when the  pump only mildly exceeds the wavebreaking threshold, the efficiency drops precipitously at larger pump intensities.

\end{abstract}


\maketitle 

\section{Introduction}


The largest laser powers are currently produced through chirped pulse amplification (CPA) technique \cite{Mourou85,Mourou98} (see also a recent review~\cite{Yakovlev_14}).  
The power limit in CPA technique comes from the final material gratings needed to re-compress the amplified pulse (which was stretched before the amplification). 
Material gratings apparently cannot tolerate laser pulses so intense that the electron quiver energy reaches the material ionization energy. 
For laser wavelengths  on the order of a micron, this limits the maximum laser intensity on gratings to a few TW$/$cm$^2$.

However, the maximum output in intensities reachable through  backward Raman amplification (BRA) of laser pulses in plasma can, in principle, be nearly 10$^6$ times larger~\cite{Malkin_99_PRL,Malkin_00_POP,Fisch_03_POP, Malkin_05_POP}.  
The BRA employs the resonant 3-wave decay of the pump laser pulse into the counter-propagating seed laser pulse  and the Langmuir wave. 
The seed pulse captures substantial fraction of the pump energy and contracts reaching nearly relativistic intensities.
Several other plasma-based mechanisms have also been proposed to compress laser pulses in a counter-propagating geometry. 
These mechanisms include Compton backscattering~\cite{Shvets_98_PRL} or, more recently, strongly-coupled Brilliouin backscattering~\cite{Weber_06,PRL-2010-Lancia,PRL-2013-Weber}, or possibly a combination of Raman and  Brilliouin backscattering~\cite{Riconda_13}.  
 However, at present, the BRA has enjoyed the most theoretical and experimental development, and appears to be the most promising  for high intensity applications. 

Inasmuch as the energy transfer in BRA is mediated by the Langmuir wave, the BRA efficiency can be significantly reduced by  Langmuir wavebreaking \cite{Malkin_99_PRL,Malkin_00_POP,Malkin_14-EPJST}, which occurs  when the longitudinal quiver electron velocity  exceeds the phase velocity of the Langmuir wave ~\cite{Dawson_59,Kruer1988}. 
Apart from the Langmuir wave breaking, the BRA efficiency might be  impeded by 
 the amplified pulse filamentation and detuning due to the relativistic electron nonlinearity \cite{Malkin_99_PRL,Fraiman_02_POP,Malkin_07_PRL,2012-dispersion,Malkin_14-EPJST,PoP-2014-Lehmann}, parasitic Raman scattering of the pump and amplified pulses by plasma noise \cite{Malkin_99_PRL,Malkin_00_PRL,Malkin_00_POP,Tsidulko_00_PRL,Solodov_04_PRE,Malkin_14-EPJST},  generation of superluminous precursors of the amplified pulse \cite{Tsidulko_02_PRL}, pulse scattering by plasma density inhomogeneities \cite{Solodov_dens}, pulse depletion and plasma heating through inverse bremsstrahlung \cite{Malkin_07_PRE,Malkin_09_PRE,Malkin_10_POP,2011-Balakin}, and resonant Langmuir wave Landau damping \cite{PRL-2005-Hur,Malkin_07_PRE,PoP-2009-Yampolsky,Malkin_10_POP,PoP-2011-Yampolsky,PoP-2012-Strozzi,IEEE-2014-Wu,NatCom-2014-Depierreux}.   Taking into account  these impediments to high efficiency, the regimes of the met robust efficiency can be identified \cite{Clark_03_POP,Yampolsky_04_PRE,Toroker_POP_12, Toroker_12_PRL}. 

In the regimes in which the wavebreaking  is not too strong, the BRA effect was demonstrated experimentally~\cite{Ping_00_PRE,Ping_02_PRE,Ping_04_PRL,Balakin_04_JETPL,
Cheng_05_PRL,Ren_08_POP,Jaro_12_NJP,Jaro_12_SPI}. 
The experiments also indicated that the maximum BRA efficiency is achieved at  pump intensities not exceeding by much the wavebreaking threshold \cite{PoP-2011-Yampolsky}, in accordance with the theoretical expectations~\cite{Malkin_99_PRL}. 

Note that, apart from the issue of efficiency, there might be advantages to operating in the parameter regime prone to  strong wavebreaking.
For example, having larger laser-to-plasma frequency ratio (at which the Langmuir phase velocity is smaller) may reduce the parasitic Raman forward scattering of the amplified pulse \cite{Malkin_99_PRL,Malkin_00_PRL,Malkin_00_POP}, while larger pump intensities might enable the amplified pulse to grow faster.
The combination of these factors can incur strong wavebreaking.

Thus,  it would be important  if there were any possibility to  increase the efficiency in strong wavebreaking regimes.
This would be primarily important around the optical range. 
For UV and X-ray regimes~\cite{Malkin_07_PRE,Malkin_09_PRE}, the wavebreaking intensities are already very high and not readily attainable at any Langmuir wave phase velocity exceeding a realistic thermal electron velocity, i.e. in the entire realistic range of the Langmuir wave existence.
Recently in PIC simulations  in the optical frequency range, high BRA
efficiency was in fact reported  in  a very strong wavebreaking regime~\cite{Trines_10}.
One of our purposes here was to confirm, in a different code, this optimistic prediction.  
However, while the efficiencies obtained here are in agreement
with most of the efficiencies reported in the recent PIC simulations,  they
do not confirm the very high efficiency in the very strong wavebreaking
regime.

Our paper explores the wavebreaking regimes numerically using the
Vlasov-Maxwell (VM) code described below.
First, we verify this code below wavebreaking.
Then we apply this code to the pump pulse intensities exceeding the
wavebreaking threshold.
For mild wavebreaking regimes, where the pump intensities that exceed the
wavebreaking threshold by no more than a factor of just several, the VM code
results  are in agreement with both analytic calculations and previous
PIC simulations.  In this regime, highly efficient backward Raman
amplification is still possible.
 For the strong wavebreaking regimes, we find that the BRA efficiency there basically agrees with both the analytical estimates of Ref.~\onlinecite{Malkin_99_PRL} and numerical results of Fig.~3a of Ref.~\onlinecite{Trines_10},  but is at variance with much higher BRA efficiency of Fig.~2a of Ref.~\onlinecite{Trines_10}. 
In addition, we show that the BRA efficiency in the mild wavebreaking regime can be noticeably increased by increasing the input seed pulse intensity, while the BRA efficiency in the strong wavebreaking regime is basically not affected by increasing the input seed pulse intensity.

\section{Model description}\label{VM_model}
To analyze the BRA wavebreaking regimes,  we employ a one-dimensional (1D) relativistic Vlasov-Maxwell (VM) code.  The non-relativistic version of this code can be found in ~\cite{Bert_90,Cheng_76,Reveille_92,Ghizzo_95}.
The VM code is applicable to the BRA both below and above the wavebreaking threshold.
In particular, below the threshold, this code covers the parameter range where the fluid description of  the BRA is applicable, while, above the threshold, this code can properly
handle kinetic effects important there.  We solve full Maxwell equations, not using an envelope approximation for waves (even though it would much reduce the computational overhead and might be particularly useful for simulating multidimensional effects~\cite{Farmer_13}), because the validity of the envelope approximations in the strong wavebreaking regime might still need to be verified independently.  

The pump and seed pulses, counter-propagating  the direction $\hat{z}$, are comprised of transverse electric and magnetic fields linearly polarized in $\hat{x}$ and $\hat{y}$  directions, respectively, $\vec{\bar E}= \bar E_x \hat{x}$ and $\vec{\bar B}= \bar B_y \hat{y} $. The seed pulse frequency $\omega_b$ is down-shifted from the pump frequency $\omega_a$ by the electron plasma frequency $\omega_e=\sqrt{4\pi n_{e0} e^2/m_e}$, so that the Langmuir wave is resonantly excited, having the longitudinal electric field $\vec{\bar E}= \bar E_z \hat{z}$. 
The fields are measured in units $m_e c \omega_e /e $,  $m_e$ is the electron mass, $-e$ is the electron charge, $n_{e0}$ is the initial electron plasma concentration and $c$ is the speed of light in vacuum. The time $\bar t$ is measured further in units $1/\omega_e$ and the distance $\bar z$ is measured in units $c/\omega_e$. 
We also define the dimensionless frequencies $\bar\omega_a=\omega_a/\omega_e$ and $\bar\omega_b=\omega_b/\omega_e$ , and the respective dimensionless wavenumbers  $\bar k_a=\sqrt{\bar\omega_a^2-1}$  and $\bar k_b=\sqrt{\bar\omega_b^2-1}$. The resonant Langmuir wave then has the dimensionless wavenumber $\bar k_f=\bar k_a+\bar k_b$.

For the fast laser-plasma interaction of interest here, the slow ion motion can be neglected.  The longitudinal electron distribution function $\bar f$ is described by the one-dimensional Vlasov equation,
\begin{eqnarray}\label{VM:vlasov}
\frac{\partial{\bar f}}{\partial{ \bar t}}+\frac{\bar p_z}{\bar \gamma}\frac{\partial \bar f}{\partial \bar z} 
- (\bar E_z + \frac{\bar P_x}{\bar \gamma}\bar B_y)\frac{\partial \bar f}{\partial \bar p_z}=0,
\end{eqnarray}
where $\bar \gamma=\sqrt{1+\bar P_x^2+\bar p_z^2}$ is the Lorentz factor, $\bar P_x$ and $\bar p_z$ are the electron momentum components in the $\hat x$ and $\hat z$ directions, respectively, measured in units $m_e c$. 
The distribution function $\bar f$ is measured in units $n_{e0}/m_e c$.

The electrostatic field $\bar E_z=-\partial \bar \phi /\partial \bar z$ is found by solving  Poisson's equation
\begin{eqnarray}\label{VM:poisson}
\frac{\partial^2 \bar \phi}{\partial \bar z^2}=-[1 -\bar n_e(\bar z,\bar t)],
\end{eqnarray}
where $\bar n_e (\bar z, \bar t )=\int \bar f(\bar z,\bar p_z, \bar t) d\bar p_z$ is the electron concentration normalized to $n_{e0}$. 

In this model,  the electron motion in $\hat{x}$ direction is described by the fluid equation
\begin{eqnarray}
\frac{\partial \bar P_x}{\partial \bar t} = - \bar E_x.
\end{eqnarray}
The electromagnetic waves are described by equations
\begin{eqnarray}\label{VM:wave_pm}
(\frac{\partial}{\partial \bar t} \pm \frac{\partial}{\partial \bar z})\bar E^{\pm} = \int d\bar p_z \frac{\bar P_x}{\bar \gamma} \bar f,
\end{eqnarray}
where $\bar E^{\pm}=\bar E_x \pm \bar B_y$.
The model Eqs.~(\ref{VM:vlasov}-\ref{VM:wave_pm}) conserves energy, 
\begin{eqnarray}\label{VM:energy}
\frac{\partial}{\partial \bar t}(W_{em} +W_{es} + W_k) = 0, 
\end{eqnarray}
where $W_{em}=\int d\bar z (\bar E_x^2+\bar B_y^2 )/2$ is the electromagnetic energy, $W_{es}=\int d\bar z \bar E_z^2/2$ is the electrostatic energy, and 
$W_k=\int d\bar z \int d \bar p_z (\bar \gamma-1)\bar f$ is the kinetic energy of the electrons.
The model presented here is similar to that of Ref.~\cite{Lehmann_13}. 

In order to avoid electromagnetic wave reflections from boundaries, perfectly matching damping layers (PML)~\cite{Bere_94,Gedney_96} are inserted  at both plasma edges.
In order to avoid the Langmuir wave reflection, a Krook~\cite{Krook_56} operator is added to the Vlasov equation that causes the electron distribution function $\bar f$ to relax to the initial distribution $\bar f_0$ in narrow boundary layers.
To exclude extra spatial length from the numerical simulations, we solve the VM equations in the window around the seed pulse, using variables 
$$\xi=\bar z + \bar t\, , \;\;\;\; \tau=\bar t\, .$$


Most of numerical examples will be presented below for  the laser-to-plasma frequency ratio $\bar\omega_b=20$ and the initial electron temperature $T_{e0}= 10\, eV$.
This temperature is much smaller than the energy of electron moving with 
the resonant Langmuir wave phase velocity, $v_{\rm ph}=\omega_e/k_f\approx c/40$, which energy is $m_e v^2_{\rm ph}/2\approx 160\, eV$.
In such a plasma, the wavebreaking occurs when the amplitude of the longitudinal electron quiver velocity, $eE_L/(m_e\omega_e)$ exceeds  $v_{\rm ph}$.  The amplitude of the Langmuir wave electric field $E_L$ at the wavebreaking threshold is then
\begin{eqnarray}\label{plasma_amp}
E_L=\frac{m_ec\,\omega_e^2}{2e\,\omega_a}\; .
\end{eqnarray}
The pump intensity at the Langmuir wavebreaking threshold can be evaluated as in~\cite{Malkin_99_PRL,Malkin_14-EPJST}.
Namely, the pump depleted energy is $\bar\omega_a \; (\approx 20)$ times larger than the energy transfered to the Langmuir wave (since decay of one pump photon produces one Langmuir plasmon of $\bar\omega_a$ smaller energy). Therefore, to produce the Langmuir wavebreaking in initially quiet plasma, the input pump intensity $I_0$ should necessarily exceed the critical wavebreaking value $I_{\rm br}$,
\begin{eqnarray}\label{plasma_energy}
I_0>I_{\rm br} =\frac{c\, \bar\omega_a}{16\pi}\left| E_L \right|^2=\frac{m_en_e c^3}{16\,\bar\omega_a} \, .
\end{eqnarray} 
For the pump of wavelength $\lambda_a=0.8$ $\mu$m and $\bar\omega_a=20$,  the wavebreaking threshold is $I_{br}=33.6$ TW$/$cm$^2$. The respective amplitude of the electron quiver velocity in the pump field is  $\bar v_{\rm br}=(2\bar\omega_a)^{-3/2}=0.004$. 


We will use the Gaussian input seed pulse of the form
\begin{eqnarray}
\bar E_{\rm seed}(\xi,\tau)=\bar v_{eb,0}\bar\omega_b\exp[-(\xi-\xi_0)^2/2\Delta_b^2].
\end{eqnarray}
In most of the examples below the input seed pulse intensity is 10 PW$/$cm$^2$,  corresponding to $\bar v_{eb,0}=0.07 $,
$\Delta_b=2\pi$ (i.e., the seed duration is one plasma period),  and $\xi_0=130$. 

The seed pulse is also characterized by the integrated seed amplitude\cite{Malkin_99_PRL, Toroker_12_PRL}
\begin{eqnarray}
U_{in}=\sqrt{\pi \bar \omega_a}\bar v_{eb,0}\Delta_b,
\end{eqnarray}
which, for the above parameters, is $U_{in}=3.5$.
 
We will calculate the relative pump depletion, $\eta$,  far enough behind the seed, at $\xi=200$,
\begin{eqnarray}
\eta=1-\frac{\left | \bar v_{ea}(\xi=200, \tau=100) \right |^2}{\left | \bar v_{ea,0} \right |^2}
\end{eqnarray} 
(here $\bar v_{ea}=\bar E_a/\bar \omega_a$ is the amplitude of the electron quiver velocity in the pump field, $\bar E_a$ is the electric field amplitude of the pump, and $\bar v_{ea,0}$ is the input value of $\bar v_{ea}$).

\section{BRA below the wavebreaking threshold}\label{BRA_bwb}
\begin{figure}
\includegraphics[width= 0.5 \textwidth]{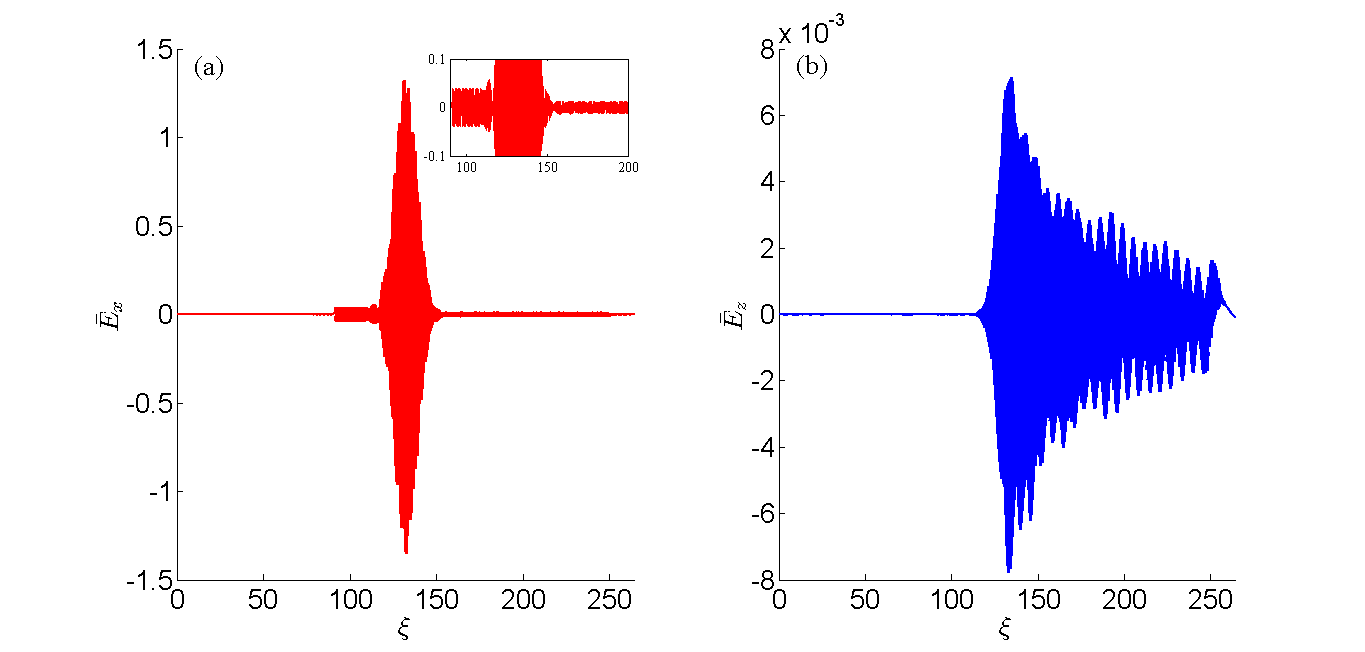}
\caption{(Color online) 
The dimensionless transverse electric field, $\bar E_x$ (Fig.~\ref{fig_1}a), and longitudinal electric field, $\bar E_z$ (Fig.~\ref{fig_1}b), at the time $\tau=140$ for the input pump intensity 4 times below the wavebreaking threshold  and the input seed pulse intensity 10 PW$/$cm$^2$.
}\label{fig_1}
\end{figure}

Consider first the well-studied case of BRA mediated by intact Langmuir wave.
Let the input pump intensity be 4 times below the wavebreaking threshold (so that $\bar v_{ea,0}=0.5 v_{\rm br})$. The pump is rectangular, injected in the positive $\xi$-direction and the front is initially located at $\xi=100$. In variables ($\xi, \tau$), the seed is not moving, while the pump propagates with the speed $2$. Figure.~\ref{fig_1} shows the transverse and longitudinal electric fields $\bar E_x$ and $\bar E_z$ at $\tau=140$.
As seen,  most of pump is depleted behind the seed pulse ($\xi>150$).

\begin{figure}
\includegraphics[width= 0.5 \textwidth]{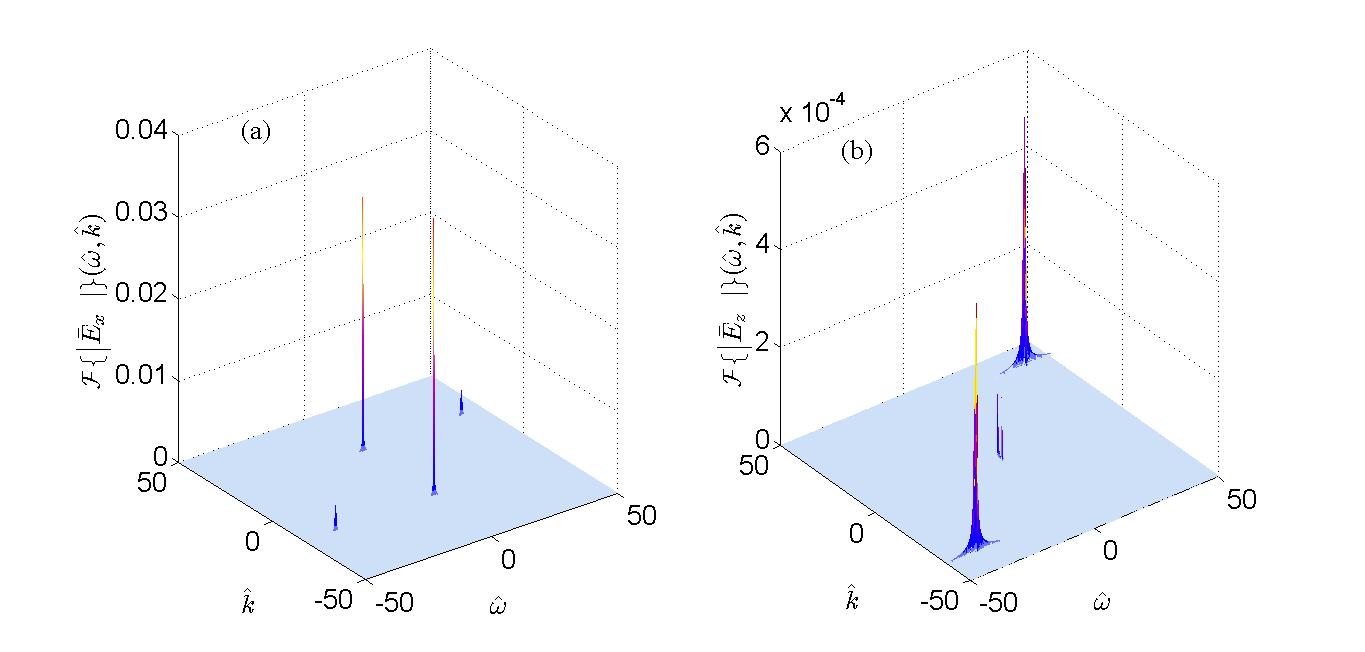}
\caption{(Color online)  
The time-space Fourier-transformed transverse,  $\bar E_x$ (Fig.~\ref{fig_2}a),  and longitudinal, $\bar E_z$ (Fig.~\ref{fig_2}b), electric fields.
}
\label{fig_2}
\end{figure}

\begin{figure}
\includegraphics[width= 0.5 \textwidth]{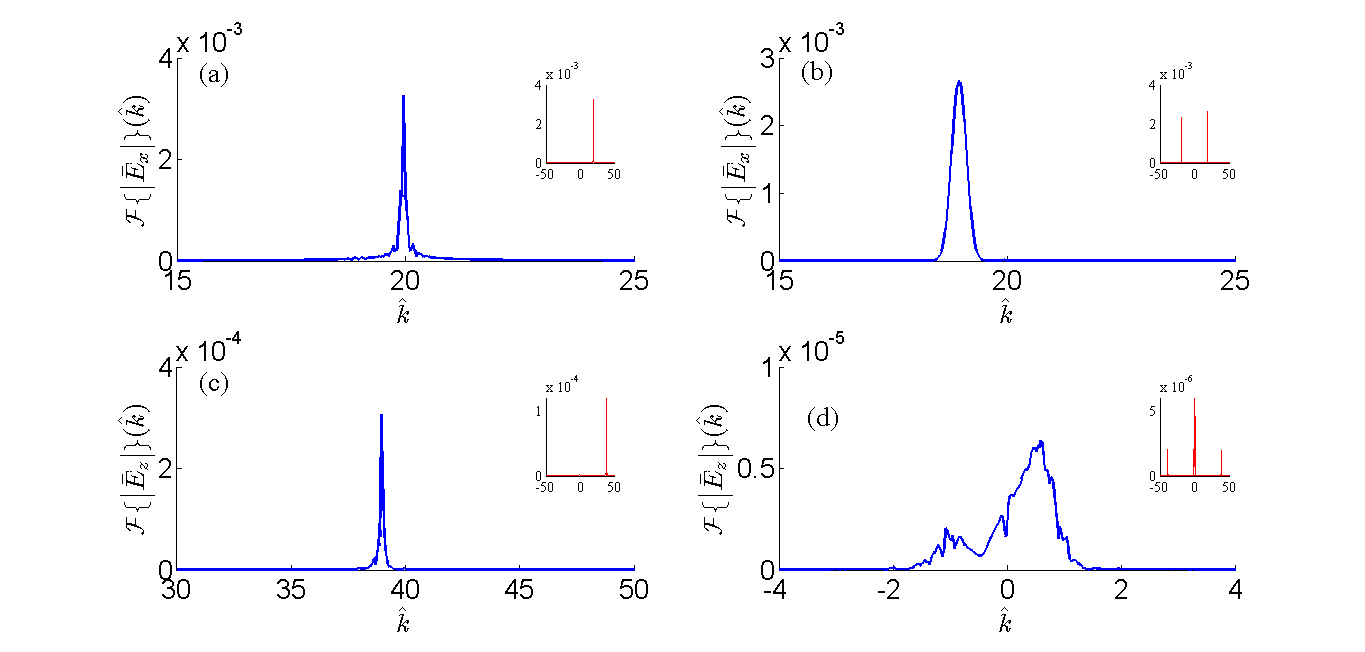}
\caption{(Color online)  
The envelope of space Fourier-transformed transverse electric field, $\bar E_x$,  at the frequency $\hat \omega = 40$ (Fig.~\ref{fig_3}a) and $\hat \omega = 0$ (Fig.~\ref{fig_3}b). The envelope of space Fourier-transformed longitudinal electric field, $\bar E_z$,  at the frequency $\hat \omega = 40$ (Fig.~\ref{fig_3}c) and $\hat \omega = 0$ (Fig.~\ref{fig_3}d).
}
\label{fig_3}
\end{figure}

To separate electromagnetic fields of different waves,we use Fourier transformation. In the variables ($\xi, \tau$), the wavenumbers are the same as in ($\bar z, \bar t$), while frequencies of the pump, seed and Langmuir wave are, respectively,
$\hat \omega_a=\bar \omega_a + \bar k_a$,  $\hat\omega_b=\bar \omega_b - \bar k_b$ and  $\hat\omega_f=\bar \omega_e + \bar k_f$. 
For  $\bar \omega_a = 20$,  the frequencies and wave numbers are
$\hat \omega_a \approx 40$, $\hat k_a \approx 20$, $\hat \omega_b \approx 0$, $\hat k_b \approx -19$, $\hat \omega_f \approx 40 $, and $\hat k_f \approx 39$.
Figure~\ref{fig_2} shows the ($\hat k$, $\hat \omega$) Fourier-transformed fields $\bar E_x$ and $\bar E_z$.
Two major spikes in the Fig.~\ref{fig_2}a for the Fourier-transformed transverse field $\bar E_x$, located at $(\hat k,\hat \omega)=(20,40)$ and $(-20,-40)$, correspond to the pump pulse,  while two lesser spikes at $(\hat k, \hat \omega)=(19,0)$ and $(-19,0)$ correspond to the seed pulse.
Fig.~\ref{fig_2}b for the Fourier-transformed longitudinal field $\bar E_z$ contains 2 major spikes, located at $(\hat k,\hat \omega)=(39,40)$ and $(\hat k,\hat \omega)=(-39,-40)$, corresponding to the resonant Langmuir wave that mediates BRA. There are also 2 lesser spikes, located at $(\hat k,\hat \omega)=(1,0)$ and $(\hat k,\hat \omega)=(-1,0)$,  corresponding to the Langmuir wave that mediates forward Raman scattering of the seed pulse.
Fig.~\ref{fig_3} shows envelopes of the spatial Fourier-transformed fields $\bar E_x$ (Figs.~\ref{fig_3}a and b) and $\bar E_z$ (Figs.~\ref{fig_3}c and d) at frequencies $\hat \omega=40$ (Figs.~\ref{fig_3}a and c) and $0$ (Figs.~\ref{fig_3}b and d). These correspond to the pump (Figs.~\ref{fig_3}a), seed (Figs.~\ref{fig_3}b) and Langmuir waves mediating BRA (Figs.~\ref{fig_3}c) and forward Raman scattering of seed pulse (Figs.~\ref{fig_3}d).

\begin{figure}
\includegraphics[width= 0.5 \textwidth]{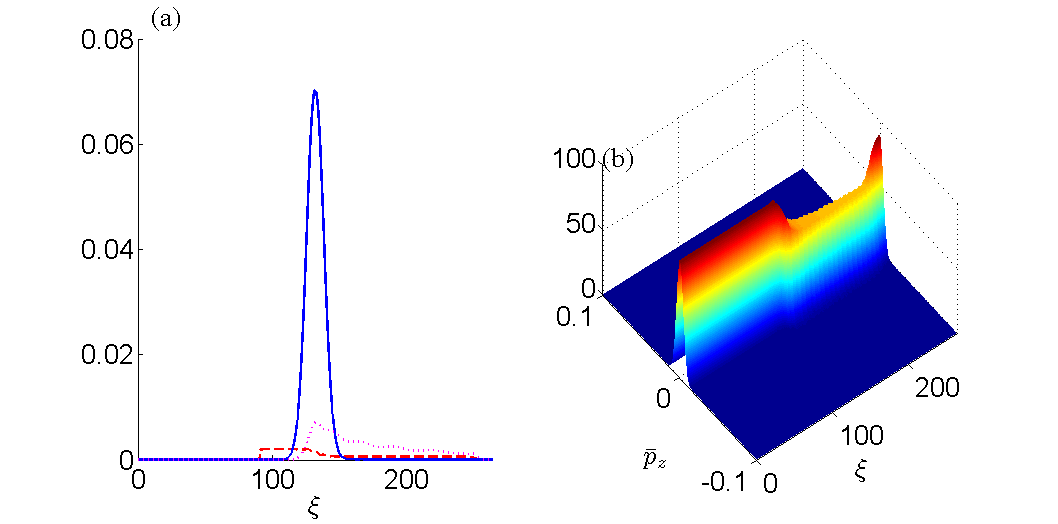}
\caption{(Color online) 
Envelopes of electron quiver velocities $\bar v_{ea}$, $\bar v_{eb}$ and $\bar v_{ef}$ in the pump (dashed line), seed (solid line) and Langmuir wave (dotted line) fields at $\tau=140$ (Fig.~\ref{fig_4}a). 
Fig.~\ref{fig_4}b shows the longitudinal electron momentum distribution function.}  
\label{fig_4}
\end{figure}

The pump, seed and Langmuir wave envelopes can be restored from the $(\hat k,\hat \omega)$ 
Fourier images using  the Hilbert transform technique~\cite{HilbTrans}.  
These envelopes are shown in Fig.~\ref{fig_4}a. The pump behind the seed pulse is depleted by $90\%$. The incomplete pump depletion can be caused by the parasitic forward Raman scattering of the seed pulse and other deleterious processes. 
Fig.~\ref{fig_4}b shows the longitudinal electron momentum  distribution function, $\bar f$, at $\tau=140$. For $\xi <130$, no interaction occurs between the pump and the seed, and the distribution function stays close to the initial Maxwellian. 
For $\xi>130$, the Langmuir wave is excited, and the distribution function is close to an oscillating Maxwellian, as it should be. 

\section{BRA in wavebreaking regimes}\label{BRA_cmp}
We'll now compare a mild wavebreaking regime, say with $\bar v_{ea}/\bar v_{br}=1.5$, i.e.,
the input pump intensity $2.25$ times above the wavebreaking threshold, to a strong wavebreaking regime with $\bar v_{ea}/\bar v_{br}=5.5$, i.e., the input pump intensity 30 times above the wavebreaking threshold.

\begin{figure}
\includegraphics[width= 0.5 \textwidth]{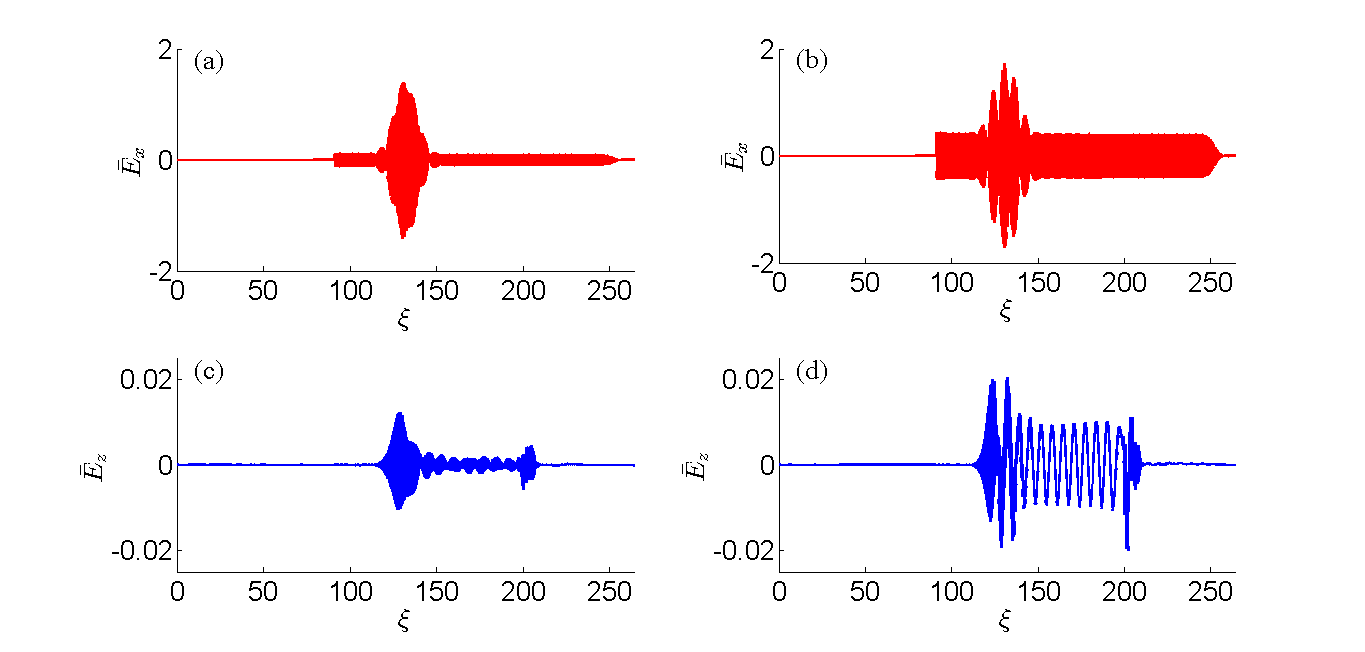}
\caption{(Color online) 
The transverse, $\bar E_x$ (solid line), and longitudinal, $\bar E_z$ (dashed line), electric fields at $\tau=90$ in the mild, $\bar v_{ea}/\bar v_{br}=1.5$ (Fig.~\ref{fig_5}a), and strong, $\bar v_{ea}/\bar v_{br}=5.5$ (Fig.~\ref{fig_5}b), wavebreaking regimes.} 
\label{fig_5}
\end{figure}

Figure~\ref{fig_5} shows the transverse and longitudinal field amplitudes in these two regimes at $\tau=90$.  Despite the wavebreaking, the longitudinal field still appears to be larger at the larger pump intensity. Nevertheless, the pump depletion behind the seed pulse drops from 30$\%$ in the mild wavebreaking regime down to 9$\%$ in the strong wavebreaking regime. 

\begin{figure}
\includegraphics[width= 0.5 \textwidth]{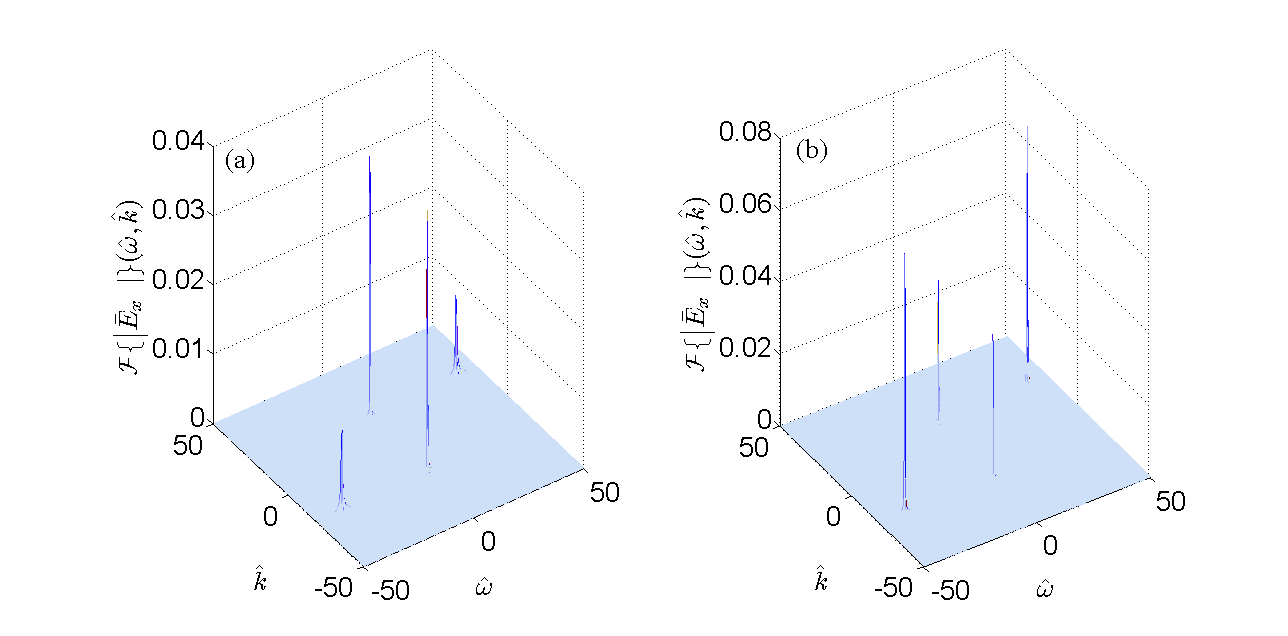}
\caption{(Color online) 
The time-space Fourier-transformed transverse field $\bar E_x$ in the mild wavebreaking regime, $\bar v_{ea}/\bar v_{br}=1.5$, (Fig.~\ref{fig_6}a)  and 
in the strong wavebreaking regime, $\bar v_{ea}/\bar v_{br}=5.5$,  (Fig.~\ref{fig_6}b). }
\label{fig_6}
\end{figure}

\begin{figure}
\includegraphics[width= 0.5 \textwidth]{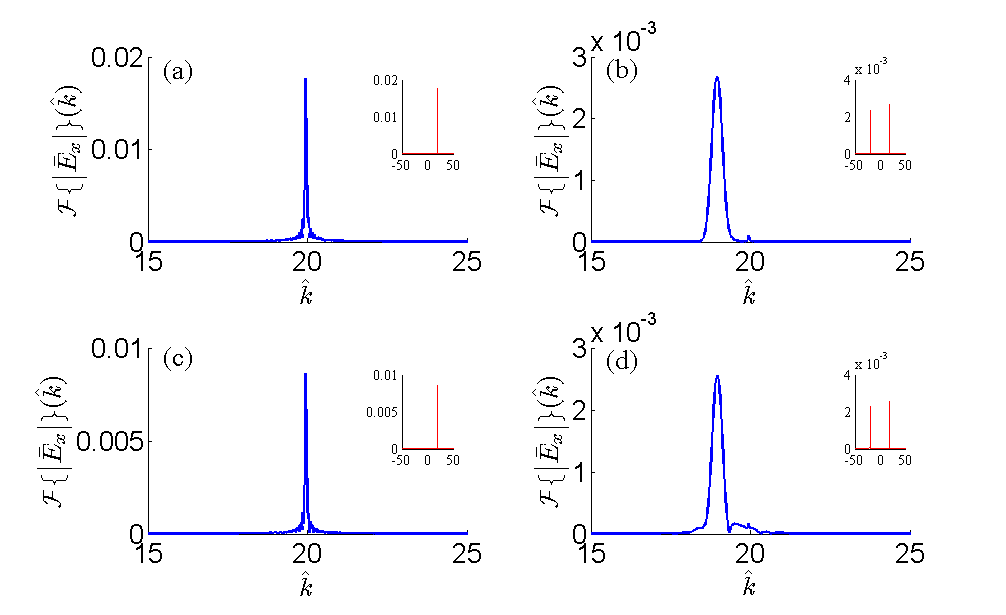}
\caption{(Color online) 
The envelope of spatially Fourier-transformed transverse field $\bar E_x$ in the mild wavebreaking regime, $\bar v_{ea}/\bar v_{br}=1.5$, at $\hat \omega = 40$ (Fig.~\ref{fig_7}a) and $\hat \omega = 0$ (Fig.~\ref{fig_7}b). 
 The envelope of spatially Fourier-transformed transverse field $\bar E_x$ in the strong wavebreaking regime, $\bar v_{ea}/\bar v_{br}=5.5$, at $\hat \omega = 40$ (Fig.~\ref{fig_7}c) and $\hat \omega = 0$ (Fig.~\ref{fig_7}d). }
\label{fig_7}
\end{figure}

Figure~\ref{fig_6} shows the Fourier-transformed transverse electric field $\bar E_x$ in these two regimes in $(\hat k,\hat \omega)$-space. 
Fig.~\ref{fig_6}a shows the mild wavebreaking case of $\bar v_{ea}/\bar v_{br}=1.5$ and Fig.~\ref{fig_6}b shows the strong wavebreaking case of $\bar v_{ea}/\bar v_{br}=5.5$.  The spatially Fourier-transformed pump and seed envelopes in the mild wavebreaking regime  are shown in Figs.~\ref{fig_7}a and b.  The spatially Fourier-transformed pump and seed envelopes in the strong wavebreaking regime are shown in Figs.~\ref{fig_7}c and d.

\begin{figure}
\includegraphics[width= 0.5 \textwidth]{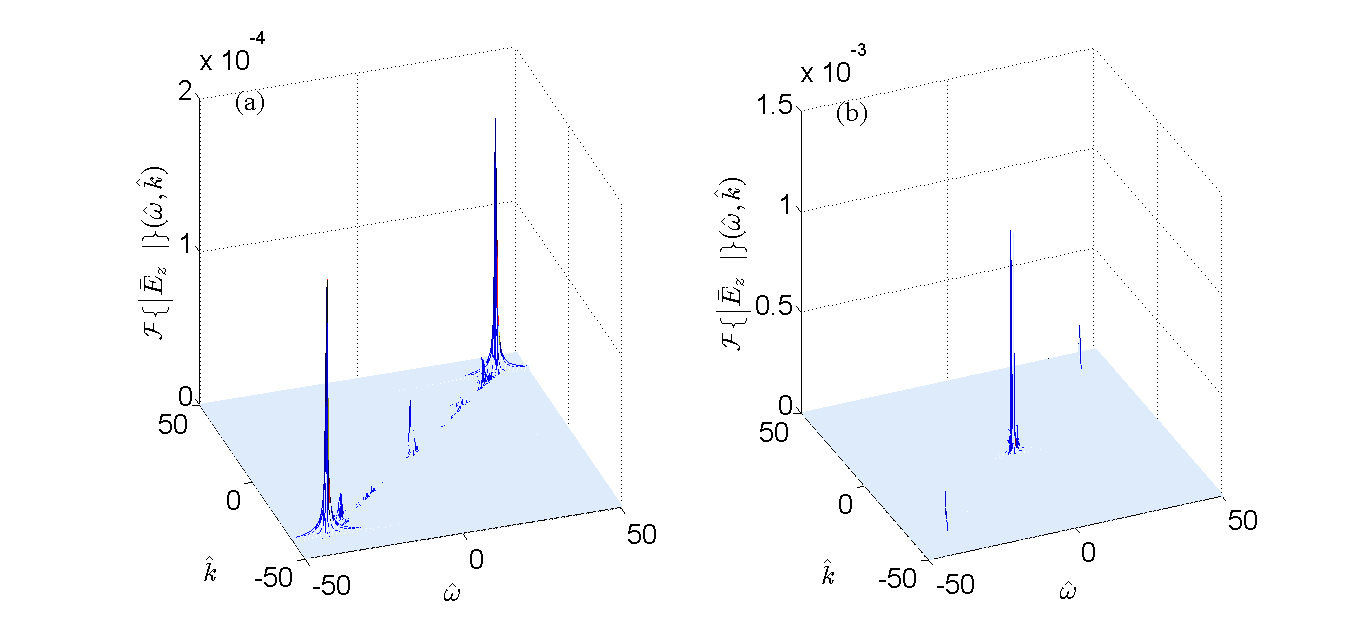}
\caption{(Color online) 
The time-space Fourier-transformed longitudinal field $\bar E_z$ in the mild, $\bar v_{ea}/\bar v_{br}=1.5$, (Fig.~\ref{fig_8}a)  and  strong, $\bar v_{ea}/\bar v_{br}=5.5$,  (Fig.~\ref{fig_8}b) wavebreaking regimes. }
\label{fig_8}
\end{figure}

\begin{figure}
\includegraphics[width= 0.5 \textwidth]{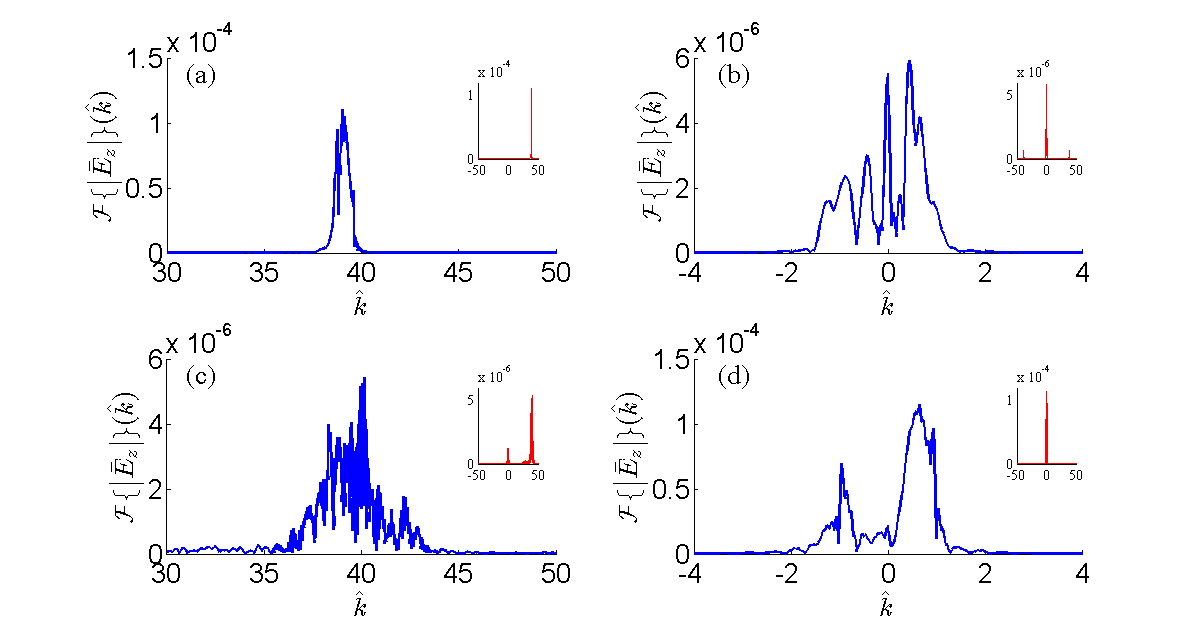}
\caption{(Color online) 
The envelope of spatially Fourier-transformed longitudinal field $\bar E_z$ in the mild wavebreaking regime, $\bar v_{ea}/\bar v_{br}=1.5$, at $\hat \omega = 40$ (Fig.~\ref{fig_9}a) and $\hat \omega = 0$ (Fig.~\ref{fig_9}b). 
The envelop of spatially Fourier-transformed longitudinal field $\bar E_z$ in the strong wavebreaking regime, $\bar v_{ea}/\bar v_{br}=5.5$, at $\hat \omega = 40$ (Fig.~\ref{fig_9}c) and $\hat \omega = 0$ (Fig.~\ref{fig_9}d).
}
\label{fig_9}
\end{figure}

Figure~\ref{fig_8} shows the Fourier-transformed longitudinal field $\bar E_z$ for the mild, $\bar v_{ea}/\bar v_{br}=1.5$ (Fig.~\ref{fig_8}a), and strong, $\bar v_{ea}/\bar v_{br}=5.5$ (Fig.~\ref{fig_8}b) wavebreaking regimes in $(\hat k,\hat \omega)$-space.
The spatially Fourier-transformed envelopes of the Langmuir waves associated with the BRA and  forward Raman scattering of the seed are shown in Fig.~\ref{fig_9}.
As seen, the bandwidth in the strong wavebreaking regime is broader than in the mild wavebreaking regime. Also, the long-wavelength components located around the spot $(\hat k,\hat \omega)=(0,0)$ are much more pronounced in the strong wavebreaking regime.

\begin{figure}
\includegraphics[width= 0.5 \textwidth]{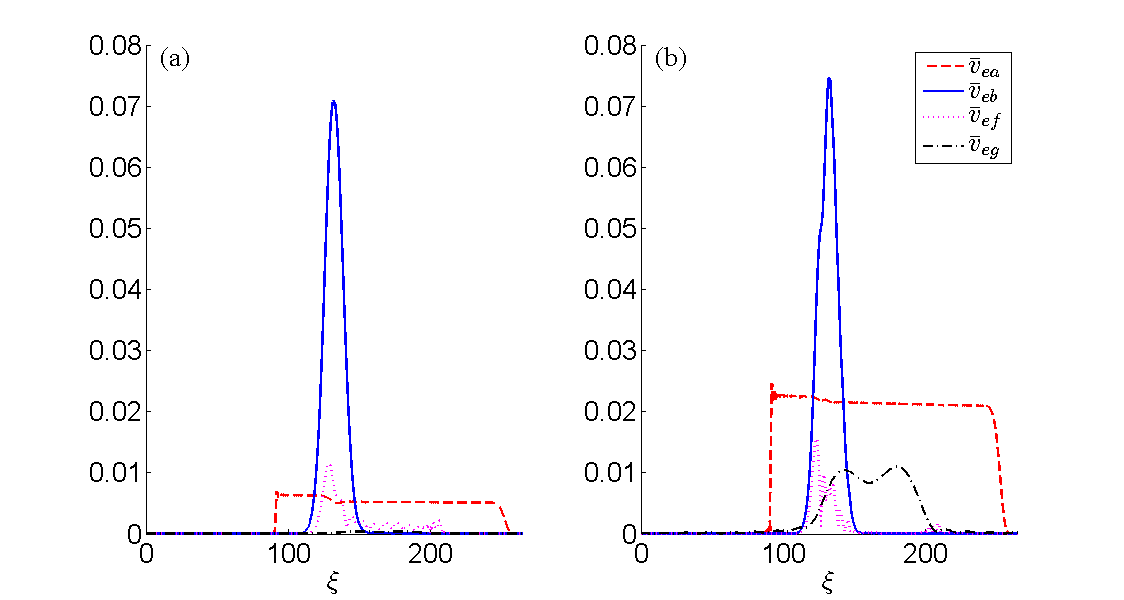}
\caption{(Color online) 
Envelopes of electron quiver velocities $\bar v_{ea}$, $\bar v_{eb}$, $\bar v_{ef}$, and $\bar v_{eg}$ in the fields of the pump pulse (dashed line), seed pulse (solid line), Langmuir wave mediating BRA (dotted line) and Langmuir wave mediating forward Raman scattering of the seed pulse (dash-dotted line) at $\tau=90$  for $\bar v_{ea}/\bar v_{br}=1.5$ (Fig.~\ref{fig_10}a)  and  $\bar v_{ea}/\bar v_{br}=5.5$ (Fig.~\ref{fig_10}b).   
}
\label{fig_10}
\end{figure}

Using $(\hat k,\hat \omega)$ Fourier images of the fields $\bar E_x$ and $\bar E_z$,  we calculated the envelopes of the pump pulse, seed pulse, and two Langmuir waves mediating BRA and forward Raman scattering of the seed pulse. Figure~\ref{fig_10} shows the results in the mild, $\bar v_{ea}/\bar v_{br}=1.5$ (Fig.~\ref{fig_10}a), and strong, $\bar v_{ea}/\bar v_{br}=5.5$ (Fig.~\ref{fig_10}b) wavebreaking regimes.

\begin{figure}
\includegraphics[width= 0.5 \textwidth]{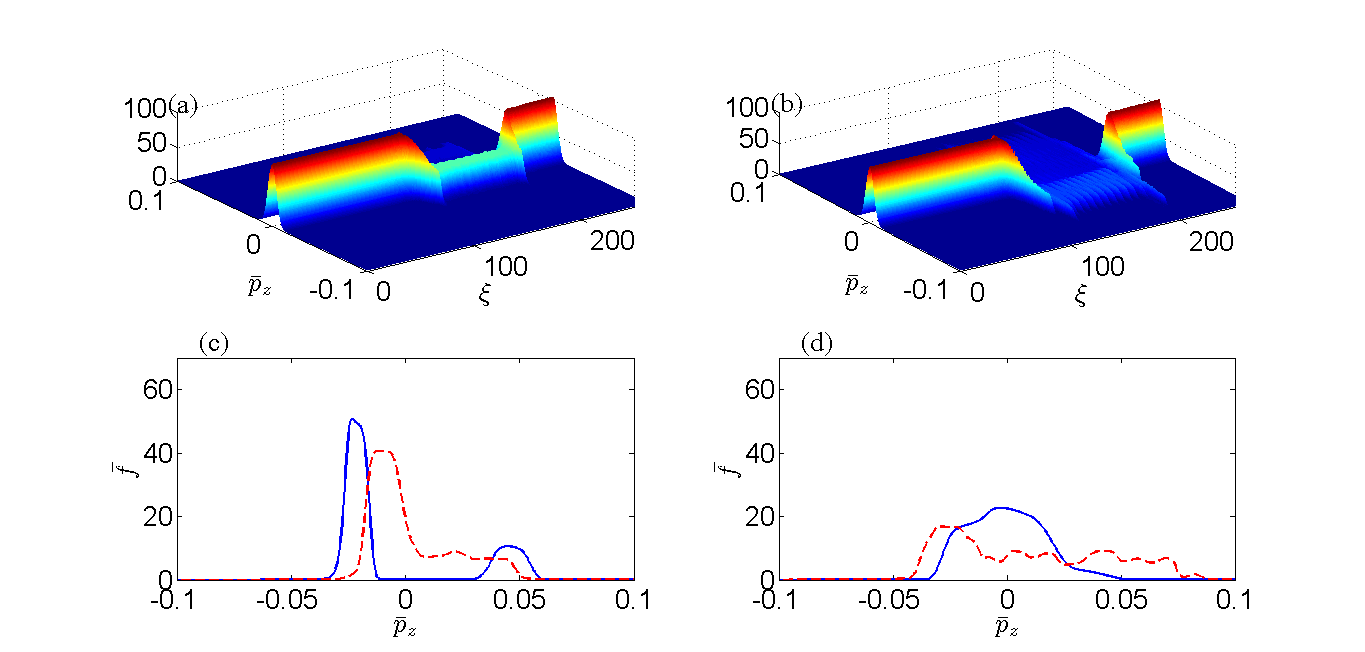}
\caption{(Color online) 
The electron distribution function for the mild, $\bar v_{ea}/\bar v_{br}=1.5$ (Figs.~\ref{fig_12}a and c), and strong,  $\bar v_{ea}/\bar v_{br}=5.5$ (Figs.~\ref{fig_12}b and d), wavebreaking regimes. Figs.~\ref{fig_12}c and d show the distribution snap-shuts at $\xi=132$ (solid line) and $\xi=150$ (dashed line). }
\label{fig_11}
\end{figure}

Fig.~\ref{fig_11} shows the electron distribution function for the mild, $\bar v_{ea}/\bar v_{br}=1.5$ (Figs.~\ref{fig_11}a and c), and strong,  $\bar v_{ea}/\bar v_{br}=5.5$ (Figs.~\ref{fig_11}b and d), wavebreaking regimes. Figs.~\ref{fig_11}c and d show the distribution snap-shuts at $\xi=132$ (solid line) and $\xi=150$ (dashed line).
The effective electron temperatures at $\xi=132$ and $\xi=150$  are $T_e=180$ eV and $470$ eV, for the mild wavebreaking regime, and $620$ eV and $870$ eV, for the strong wavebreaking regime, respectively.

\begin{figure}
\includegraphics[width= 0.5 \textwidth]{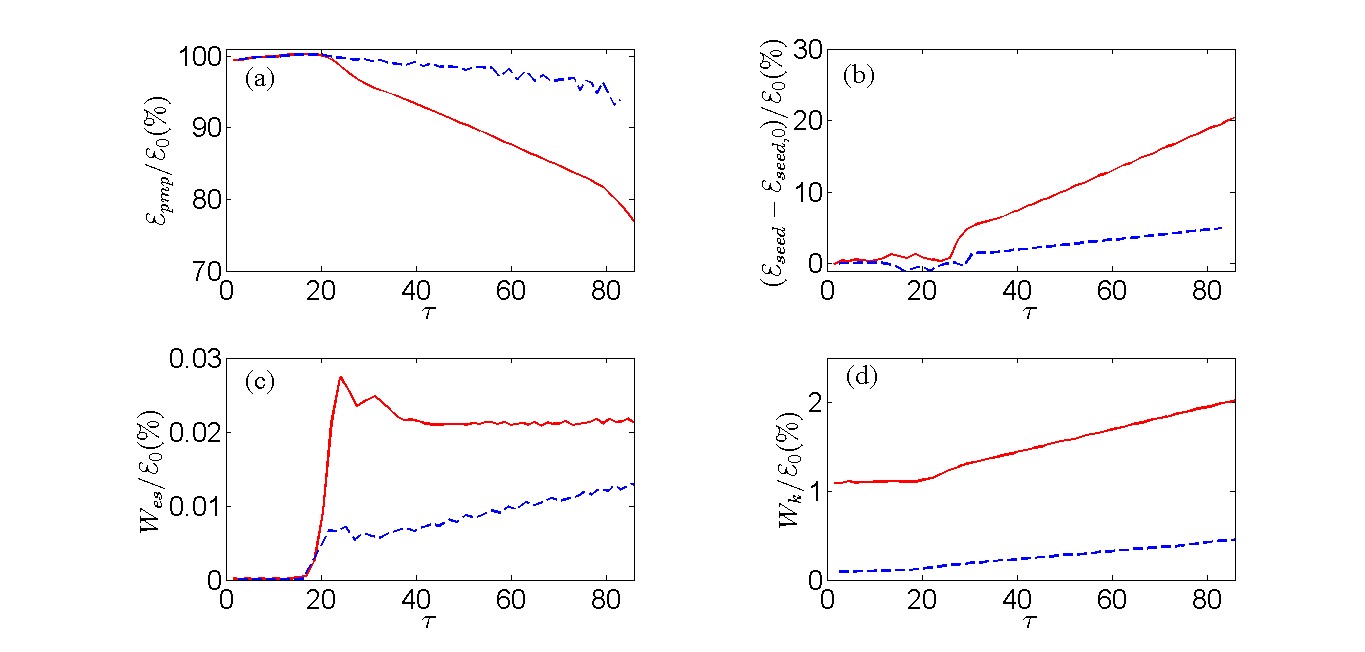}
\caption{(Color online) 
Percentage of input pump energy remaining in the pump (Fig.~\ref{fig_12}a), transferred to the seed (Fig.~\ref{fig_12}b), electrostatic field (Fig.~\ref{fig_12}c), and  plasma electrons (Fig.~\ref{fig_12}d)  in the mild, $\bar v_{ea}/\bar v_{br}=1.5$ (solid curve) and strong, $\bar v_{ea}/\bar v_{br}=5.5$ (dashed curve) wavebreaking regimes.
 }
\label{fig_12}
\end{figure}

Finally, Fig.~\ref{fig_12} shows  the fraction of pump energy that is decayed (a), transferred to the seed (b), to electrostatic waves (c), and to plasma electrons (d), in the mild (solid curve) and strong  (dashed curve) wavebreaking regimes. As seen, the pump depletion is significantly larger  in the mild wavebreaking regime.

\section{Discussion}

The results obtained here are by and large in agreement with previously reported PIC simulations.  However there are also significant discrepancies.  In this section, the VM simulations presented here are compared both to previous PIC simulations as well as to theoretical expectations.

\begin{figure}
\includegraphics[width= 0.5 \textwidth]{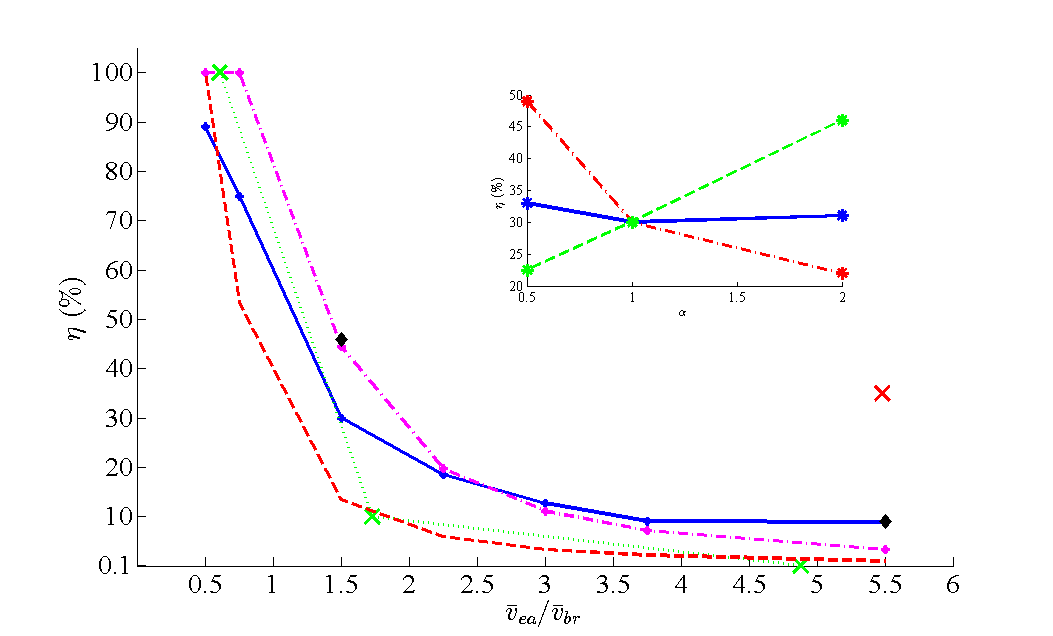}
\caption{(Color online) The pump depletion calculated numerically using VM code for the initial electron temperature $T_e=10$ eV and input seed intensity 10 PW$/$cm$^2$ (solid line), the analytical estimate of the Ref. ~\onlinecite{Malkin_99_PRL}, $\eta\sim(\bar v_{br}/\bar v_{ea})^2$, (dash-dot line), the same estimate with a smaller numerical coefficient, $\eta\sim 0.3\times(\bar v_{br}/\bar v_{ea})^2$ (dashed line);
the pump depletion reported in PIC simulations Ref.~\onlinecite{Trines_10}, Fig. 3a, (crosses at 
$\bar v_{br}/\bar v_{ea}=0.61$, $1.73$, and $4.88$) and in the same Ref.~\onlinecite{Trines_10}, Fig.2a (cross at $\bar v_{br}/\bar v_{ea}=5.5$).
The diamonds show our VM results at larger input seed intensities,  40 PW$/$cm$^2$ (the diamond at $\bar v_{ea}/\bar v_{br}=1.5$) and 100 PW$/$cm$^2$ (the diamond at $\bar v_{ea}/\bar v_{br}=5.5$).
The inset shows how the pump depletion depends on the input seed duration and amplitude:
the solid line corresponds to constant seed amplitude, the dash-dot line corresponds to constant seed capacity, and the dashed line corresponds to constant seed duration.}
\label{fig_13}
\end{figure}

This comparison is made in  Fig.~\ref{fig_13}, which shows  the relative pump depletion, $\eta$, calculated using our VM code to the results of PIC code simulations ~\cite{Trines_10}, as well as to the analytical estimate~\cite{Malkin_99_PRL} for the strong wavebreaking regime $\bar v_{ea}/\bar v_{br} \gg 1$.
The solid line is based on our VM simulations at the initial electron temperature $T_e=10$ eV and input seed intensity 10 PW$/$cm$^2$.
 The dash-dotted line shows the analytical estimate~\cite{Malkin_99_PRL}, $\eta\sim(\bar v_{br}/\bar v_{ea})^2$.
 The  dashed line is the same estimate with a smaller numerical coefficient, $\eta\sim 0.3\times(\bar v_{br}/\bar v_{ea})^2$.
 The crosses at 
$\bar v_{br}/\bar v_{ea}=0.61$, $1.73$, and $4.88$ show the pump depletion calculated through PIC simulations and reported in Fig.~3a of Ref.~\onlinecite{Trines_10}.
The cross at $\bar v_{br}/\bar v_{ea}=5.5$ shows the pump depletion reported in Fig.~2a of the same Ref.~\onlinecite{Trines_10}. 
Finally, the diamonds show our VM results at larger input seed intensities,  40 PW$/$cm$^2$ (the diamond at $\bar v_{ea}/\bar v_{br}=1.5$) and 100 PW$/$cm$^2$ (the diamond at $\bar v_{ea}/\bar v_{br}=5.5$) .

It can be seen that the PIC results presented in  Fig.~3a of Ref.~\onlinecite{Trines_10}  agree reasonably well both with the analytical estimate of Ref.~\onlinecite{Malkin_99_PRL} and with our VM simulations. 
There is somewhat smaller pump depletion in Fig.~3a of Ref.~\onlinecite{Trines_10}.
This might be due to premature backscattering of the pump by  PIC noise in the simulations of Ref.~\onlinecite{Trines_10}. 
Note that numerical noise would act much like physical noise in inducing premature backscattering.
Note also that, although not employed in these simulations, the premature  backscattering of the pump by noise, whether physical or numerical noise, could, in principle, be suppressed by  selective resonance detuning techniques~\cite{Malkin_00_PRL,Malkin_00_POP,Malkin_14-EPJST}. In any event, there is not large  discrepancy between  results presented in Fig.~3a of Ref.~\onlinecite{Trines_10} and both our VM  simulations and with the analytical estimate.


The large discrepancy occurs in the strong wavebreaking regime case shown in Fig.~2a of Ref. \onlinecite{Trines_10}, where the  pump intensity is 30 times higher than the wavebreaking threshold ($\bar v_{ea}/\bar v_{br}=5.5$).
Here, the PIC results report a surprisingly high  $35\%$ BRA efficiency. 
This high efficiency disagrees with both our VM numerical simulations results and the analytical estimate of Ref.~\onlinecite{Malkin_99_PRL}.
The high efficiency also appear even to disagree with the efficiency shown  in Fig.~3a of the same Ref. \onlinecite{Trines_10}.
In contrast to the 35\% efficiency,   the VM simulation, the analytical estimate, and Fig.~3a all  give less than $10\%$ pump depletion for such a regime.


The inset of Fig.~\ref{fig_13} shows how the pump depletion depends on the input seed duration and amplitude in the mild wavebreaking regime, $\bar v_{ea}/\bar v_{br}=1.5$.
Results for the constant initial seed duration of one plasma period and few input seed amplitudes, marked in the inset of Fig.~\ref{fig_13} by few different values of the parameter $\alpha=\bar v_{eb}/\bar v_{eb,0}$, are shown by the dashed line.
Results for  constant input seed amplitude, $\bar v_{eb}= 0.07 $, and few input seed durations, marked by few values the parameter  $\alpha=\bar \Delta_{b}/\bar \Delta_{b,0}$, are shown by the solid line. Results for  constant seed integrated amplitude $U_{in}=3.5$ and few input seed durations  are shown by the dash-dot line. Our results from the inset indicate that in the mild wavebreaking regime it is beneficial to choose high initial seed pulse intensity to obtain maximal BRA efficiency.

\section{Summary}\label{Concl}
The wavebreaking BRA regime in strongly undercritical plasma ($\omega_a/\omega_e=20$) was studied using a 1D Maxwell-Vlasov code.
This code confirmed that efficient BRA is possible for the pump pulse intensities up to a few times larger than the wavebreaking threshold. 
However, for  pump intensities exceeding by more than a factor of 10 the wavebreaking threshold, the amplification efficiency significantly decreases. 

For example, for the pump intensity exceeding the wavebreaking threshold by a factor of 30, we only found possible a BRA efficiency of  less than $10\%$. 
This low efficiency is consistent both with the analytical estimate of Ref. \onlinecite{Malkin_99_PRL} and with  Fig.~3a of Ref. \onlinecite{Trines_10}.
 However, this low efficiency is at variance with  Fig.~2a of the same Ref. \onlinecite{Trines_10},  where the rather higher efficiency of $35\%$  was reported. 
 It remains of interest, but reserved for a future study, to consider why in fact this difference is so large.

A further important  finding of this study is that, in the strong wavebreaking regime, in contrast to the mild wavebreaking regime, increasing the seed pulse intensity does not increase the BRA efficiency.


\begin{acknowledgments}
This work was supported by the NNSA SSAA under grant number  DE274-FG52- 08NA28553 and by the NSF under Grant PHY-1202162.

\end{acknowledgments}
%

\begin{thebibliography}{10}
\newcommand{\enquote}[1]{``#1''}
\expandafter\ifx\csname url\endcsname\relax
  \def\url#1{{#1}}\fi
\expandafter\ifx\csname urlprefix\endcsname\relax\def\urlprefix{}\fi

\bibitem{Mourou85}
D.~Strickland and G.~Mourou, \enquote{Compression of amplified chirped optical
  pulses,} Opt.~Commun. {\bf 56}, 219 (1985).

\bibitem{Mourou98}
G.~A. Mourou, C.~P.~J. Barty, and M.~D. Perry, \enquote{Ultrahigh-intensity
  lasers: physics of the extreme on a tabletop,} Phys.~Today {\bf 51}, 22
  (1998).

\bibitem{Yakovlev_14}
I.~V. Yakovlev, \enquote{Stretchers and compressors for ultra-high power laser
  systems,} Quantum Electronics {\bf 44}, 393 (2014).

\bibitem{Malkin_99_PRL}
V.~M. Malkin, G.~Shvets, and N.~J. Fisch, \enquote{Fast compression of laser
  beams to highly overcritical powers,} Phys. Rev. Lett. {\bf 82}, 4448 (1999).

\bibitem{Malkin_00_POP}
V.~M. Malkin, G.~Shvets, and N.~J. Fisch, \enquote{Ultra-powerful compact
  amplifiers for short laser pulses,} Phys. Plasmas {\bf 7}, 2232 (2000).

\bibitem{Fisch_03_POP}
N.~J. Fisch and V.~M. Malkin, \enquote{Generation of ultrahigh intensity laser
  pulses,} Phys. Plasmas {\bf 10}, 2056 (2003).

\bibitem{Malkin_05_POP}
V.~M. Malkin and N.~J. Fisch, \enquote{Manipulating ultra-intense laser pulses
  in plasmas,} Phys. Plasmas {\bf 12}, 044\,507 (2005).

\bibitem{Shvets_98_PRL}
G.~Shvets, N.~J. Fisch, A.~Pukhov, and J.~{Meyer-ter-Vehn},
  \enquote{Supperradiant amplification of an ultrashort laser pulse in a plasma
  by a counterpropagating pump,} Phys. Rev. Lett. {\bf 81}, 4879 (1998).

\bibitem{Weber_06}
A.~A. Andreev, C.~Riconda, V.~T. Tikhonchuk, and S.~Weber, \enquote{Short light
  pulse amplification and compression by stimulated Brillouin scattering in
  plasmas in the strong coupling regime,} Physics of Plasmas {\bf 13}, 053\,110
  (2006).

\bibitem{PRL-2010-Lancia}
L.~Lancia, J.-R. Marqu\`es, M.~Nakatsutsumi, C.~Riconda, S.~Weber, S.~H\"uller,
  A.~Man\ifmmode \check{c}\else \v{c}\fi{}i\ifmmode~\acute{c}\else \'{c}\fi{},
  P.~Antici, V.~T. Tikhonchuk, A.~H\'eron, P.~Audebert, and J.~Fuchs,
  \enquote{Experimental Evidence of Short Light Pulse Amplification Using
  Strong-Coupling Stimulated Brillouin Scattering in the Pump Depletion
  Regime,} Phys. Rev. Lett. {\bf 104}, 025\,001 (2010).

\bibitem{PRL-2013-Weber}
S.~Weber, C.~Riconda, L.~Lancia, J.-R. Marqu\`es, G.~A. Mourou, and J.~Fuchs,
  \enquote{Amplification of Ultrashort Laser Pulses by Brillouin Backscattering
  in Plasmas,} Phys. Rev. Lett. {\bf 111}, 055\,004 (2013).

\bibitem{Riconda_13}
C.~Riconda, S.~Weber, L.~Lancia, J.~Marques, G.~A. Mourou, and J.~Fuchs,
  \enquote{Spectral characteristics of ultra-short laser pulses in plasma
  amplifiers,} Physics of Plasmas {\bf 20}, 083\,115 (2013).

\bibitem{Malkin_14-EPJST}
V.~M. Malkin and N.~J. Fisch, \enquote{Key plasma parameters for resonant
  backward Raman amplification in plasma,} Eur. Phys. J. Special Topics {\bf
  223}, 1157 (2014).

\bibitem{Dawson_59}
J.~M. Dawson, \enquote{Nonlinear Electron Oscillations in a Cold Plasma,} Phys.
  Rev. {\bf 113}, 383 (1959).

\bibitem{Kruer1988}
W.~L. Kruer, {\em The Physics of Laser Plasma Interactions\/} (Addison-Wesley,
  Reading, MA, 1988).

\bibitem{Fraiman_02_POP}
G.~M. Fraiman, N.~A. Yampolsky, V.~M. Malkin, and N.~J. Fisch,
  \enquote{Robustness of laser phase fronts in backward Raman amplifiers,}
  Phys. Plasmas {\bf 9}, 3617 (2002).

\bibitem{Malkin_07_PRL}
V.~M. Malkin and N.~J. Fisch, \enquote{Relic crystal-lattice effects on Raman
  compression of powerful x-ray pulses in plasmas,} Phys. Rev. Lett. {\bf 99},
  205\,001 (2007).

\bibitem{2012-dispersion}
V.~M. Malkin, Z.~Toroker, and N.~J. Fisch, \enquote{Laser duration and
  intensity limits in plasma backward Raman amplifiers,} Phys. Plasmas {\bf
  19}, 023\,109 (2012).

\bibitem{PoP-2014-Lehmann}
G.~Lehmann and K.~H. Spatschek, \enquote{{Non-filamentated ultra-intense and
  ultra-short pulse fronts in three-dimensional Raman seed amplification},}
  Phys. Plasmas {\bf {21}}, 053\,101 ({2014}).

\bibitem{Malkin_00_PRL}
V.~M. Malkin, G.~Shvets, and N.~J. Fisch, \enquote{Detuned Raman amplification
  of short laser pulses in plasma,} Phys. Rev. Lett. {\bf 84}, 1208 (2000).

\bibitem{Tsidulko_00_PRL}
V.~M. Malkin, Y.~A. Tsidulko, and N.~J. Fisch, \enquote{Stimulated Raman
  scattering of rapidly amplified short laser pulses,} Phys. Rev. Lett. {\bf
  85}, 4068 (2000).

\bibitem{Solodov_04_PRE}
A.~A. Solodov, V.~M. Malkin, and N.~J. Fisch, \enquote{Pump side scattering in
  ultrapowerful backward Raman amplifiers,} Phys. Rev. E {\bf 69}, 066\,413
  (2004).

\bibitem{Tsidulko_02_PRL}
Y.~A. Tsidulko, V.~M. Malkin, and N.~J. Fisch, \enquote{Suppression of
  superluminous precursors in high-power backward Raman amplifiers,} Phys. Rev.
  Lett. {\bf 88}, 235\,004 (2002).

\bibitem{Solodov_dens}
A.~A. Solodov, V.~M. Malkin, and N.~J. Fisch, \enquote{Random density
  inhomogeneities and focusability of the output pulses for plasma-based
  powerful backward Raman amplifiers,} Phys. Plasmas {\bf 10}, 2540 (2003).

\bibitem{Malkin_07_PRE}
V.~M. Malkin, N.~J. Fisch, and J.~S. Wurtele, \enquote{Compression of powerful
  x-ray pulses to attosecond durations by stimulated Raman backscattering in
  plasmas,} Phys. Rev. E {\bf 75}, 026\,404 (2007).

\bibitem{Malkin_09_PRE}
V.~M. Malkin and N.~J. Fisch, \enquote{Quasitransient regimes of backward Raman
  amplification of intense x-ray pulses,} Phys. Rev. E {\bf 80}, 046\,409
  (2009).

\bibitem{Malkin_10_POP}
V.~M. Malkin and N.~J. Fisch, \enquote{Quasitransient backward Raman
  amplification of powerful laser pulses in plasma with multicharged ions,}
  Phys. Plasmas {\bf 17}, 073\,109 (2010).

\bibitem{2011-Balakin}
A.~A. Balakin, N.~J. Fisch, G.~M. Fraiman, V.~M. Malkin, and Z.~Toroker,
  \enquote{Numerical modeling of quasitransient backward Raman amplification of
  laser pulses in moderately undercritical plasmas with multicharged ions,}
  Phys. Plasmas {\bf 18}, 102\,311 (2011).

\bibitem{PRL-2005-Hur}
M.~S. Hur, R.~R. Lindberg, A.~E. Charman, J.~S. Wurtele, and H.~Suk,
  \enquote{Electron Kinetic Effects on Raman Backscatter in Plasmas,} Phys.
  Rev. Lett. {\bf 95}, 115\,003 (2005).

\bibitem{PoP-2009-Yampolsky}
N.~Yampolsky and N.~Fisch, \enquote{Effect of nonlinear Landau damping in
  plasma-based backward Raman amplifier,} Phys. Plasmas {\bf 16}, 072\,105
  (2009).

\bibitem{PoP-2011-Yampolsky}
N.~Yampolsky and N.~Fisch, \enquote{Limiting effects on laser compression by
  resonant backward Raman scattering in modern experiments,} Physics of Plasmas
  {\bf 18}, 056\,711 (2011).

\bibitem{PoP-2012-Strozzi}
D.~Strozzi, E.~Williams, H.~Rose, D.~Hinkel, A.~Langdon, and J.~Banks,
  \enquote{Threshold for electron trapping nonlinearity in Langmuir waves,}
  Phys. Plasmas {\bf 19}, 112\,306 (2012).

\bibitem{IEEE-2014-Wu}
Z.~Wu, Y.~Zuo, J.~Su, L.~Liu, Z.~Zhang, and X.~Wei, \enquote{Production of
  single pulse by Landau damping for backward Raman amplification in plasma,}
  IEEE Transactions on Plasma Science {\bf 42}, 1704--1708 (2014).

\bibitem{NatCom-2014-Depierreux}
S.~Depierreux, V.~Yahia, C.~Goyon, G.~Loisel, P.-E. Masson-Laborde,
  N.~Borisenko, A.~Orekhov, O.~Rosmej, T.~Rienecker, and C.~Labaune,
  \enquote{Laser light triggers increased Raman amplification in the regime of
  nonlinear Landau damping,} Nature Communications {\bf 5}, 4158 (2014).

\bibitem{Clark_03_POP}
D.~S. Clark and N.~J. Fisch, \enquote{Operating regime for a backward Raman
  laser amplifier in preformed plasma,} Phys. Plasmas {\bf 10}, 3363 (2003).

\bibitem{Yampolsky_04_PRE}
N.~A. Yampolsky, V.~M. Malkin, and N.~J. Fisch, \enquote{Finite-duration
  seeding effects in powerful backward Raman amplifiers,} Phys. Rev. E {\bf
  69}, 036\,401 (2004).

\bibitem{Toroker_POP_12}
Z.~Toroker, V.~M. Malkin, A.~A. Balakin, G.~M. Fraiman, and N.~J. Fisch,
  \enquote{Geometrical constraints on plasma couplers for Raman compression,}
  Phys.~Plasmas {\bf 19}, 083\,110 (2012).

\bibitem{Toroker_12_PRL}
Z.~Toroker, V.~M. Malkin, and N.~J. Fisch, \enquote{Seed Laser Chirping for
  Enhanced Backward Raman Amplification in Plasmas,} Phys. Rev. Lett. {\bf
  109}, 085\,003 (2012).

\bibitem{Ping_00_PRE}
Y.~Ping, I.~Geltner, N.~J. Fisch, G.~Shvets, and S.~Suckewer,
  \enquote{Demonstration of ultrashort laser pulse amplification in plasmas by
  a counterpropagating pumping beam,} Phys. Rev. E {\bf 62}, R4532 (2000).

\bibitem{Ping_02_PRE}
Y.~Ping, I.~Geltner, A.~Morozov, N.~J. Fisch, and S.~Suckewer, \enquote{Raman
  amplification of ultrashort laser pulses in microcapillary plasmas,} Phys.
  Rev. E {\bf 66}, 046\,401 (2002).

\bibitem{Ping_04_PRL}
Y.~Ping, W.~Cheng, S.~Suckewer, D.~S. Clark, and N.~J. Fisch,
  \enquote{Amplification of ultrashort laser pulses by a resonant Raman scheme
  in a gas-jet plasma,} Phys. Rev. Lett. {\bf 92}, 175\,007 (2004).

\bibitem{Balakin_04_JETPL}
A.~A. Balakin, D.~V. Kartashov, A.~M. Kiselev, S.~A. Skobelev, A.~N. Stepanov,
  and G.~M. Fraiman, \enquote{Laser pulse amplification upon Raman
  backscattering in plasma produced in dielectric capillaries,} JETP Lett. {\bf
  80}, 12 (2004).

\bibitem{Cheng_05_PRL}
W.~Cheng, Y.~Avitzour, Y.~Ping, S.~Suckewer, N.~J. Fisch, M.~S. Hur, and J.~S.
  Wurtele, \enquote{Reaching the nonlinear regime of Raman amplification of
  ultrashort laser pulses,} Phys. Rev. Lett. {\bf 94}, 045\,003 (2005).

\bibitem{Ren_08_POP}
J.~Ren, S.~Li, A.~Morozov, S.~Suckewer, N.~A. Yampolsky, V.~M. Malkin, and
  N.~J. Fisch, \enquote{A compact double-pass Raman backscattering
  amplifier/compressor,} Phys. Plasmas {\bf 15}, 056\,702 (2008).

\bibitem{Jaro_12_NJP}
G.~Vieux, A.~Lyachev, X.~Yang, B.~Ersfeld, J.~P. Farmer, E.~Brunetti, R.~C.
  Issac, G.~Raj, G.~H. Welsh, S.~M. Wiggins, and D.~A. Jaroszynski,
  \enquote{Chirped pulse Raman amplification in plasma,} New J. Phys. {\bf 13},
  063\,042 (2011).

\bibitem{Jaro_12_SPI}
X.~Yang, G.~Vieux, E.~Brunetti, J.~P. Farmer, B.~Ersfeld, S.~M. Wiggins, R.~C.
  Issac, G.~H. Welsh, and D.~A. Jaroszynski, \enquote{Experimental
  investigation of chirp pulse Raman amplification in plasma,} Proc. SPIE {\bf
  8075}, 80\,750G (2011).

\bibitem{Trines_10}
R.~M.~G.~M. Trines, F.~Fiuza, R.~Bingham, R.~A. Fonseca, L.~O. Silva, R.~A.
  Cairns, and P.~A. Norreys, Nature Phys. {\bf 7}, 87 (2011).

\bibitem{Bert_90}
P.~Bertrand, A.~Ghizzo, T.~W. Johnston, M.~Shoucri, E.~Fijalkow, and M.~R.
  Feix, \enquote{A nonperiodic Euler-Vlasov code for the numerical simulation
  of laser-plasma beat wave acceleration and Raman scattering,} Phys. Fluids B
  {\bf 2}, 1028 (1990).

\bibitem{Cheng_76}
C.~Z. Cheng and G.~Knorr, \enquote{The integration of the Vlasov Equation in
  Configuration Space,} J. Comp. Phys. {\bf 22}, 330 (1976).

\bibitem{Reveille_92}
T.~Reveille, P.~Bertrand, A.~Ghizzo, J.~Lebas, T.~W. Johnston, and M.~Shoucri,
  \enquote{Stimulated Raman scattering: Close correspondence of Vlasov
  simulation and coupled modes,} Phys. Fluids B {\bf 4}, 2665 (1992).

\bibitem{Ghizzo_95}
A.~Ghizzo, T.~Reveille, P.~Bertrand, T.~Johntson, J.~Lebas, and M.~Shoucri,
  \enquote{An Eulerian Vlasov-Hilbert Code for the Numerical-Simulation of the
  Interaction of High-Frequency Electromagnetic-Waves with Plasma,} J. Comp.
  Phys. {\bf 118}, 356--365 (1995).

\bibitem{Farmer_13}
J.~P. Farmer and A.~Pukhov, \enquote{Fast multidimensional model for the
  simulation of Raman amplification in plasma,} Phys.~Rev. E {\bf 88}, 063\,104
  (2013).

\bibitem{Lehmann_13}
G.~Lehmann, K.~H. Spatschek, and G.~Sewell, \enquote{Pulse shaping during
  Raman-seed amplification for short laser pulses,} Phys.~Rev.~E {\bf 87},
  063\,107 (2013).



\bibitem{Bere_94}
J.-P. Berenger, \enquote{A Perfectly Matched Layer for the Absorption of
  Electromagnetic Waves,} J. Comp. Phys. {\bf 114}, 185--200 (1994).

\bibitem{Gedney_96}
S.~D. Gedney, \enquote{An Anisotropic Perefectly Matched Layer-Absorbing Medium
  for the Truncation of FDTD Lattices,} IEEE Trans. Antennas Propag. {\bf 44},
  1630 (1996).

\bibitem{Krook_56}
E.~P. Gross and M.~Krook, \enquote{Model for Collision Processes in Gases:
  Small-Amplitude Oscillations of Charged Two-Component Systems,} Phys. Rev.
  {\bf 102}, 593--604 (1956).

\bibitem{HilbTrans}
E.~Bedrosian, \enquote{A Product Theorem for Hilbert Transforms,} Rand
  Corporation Memorandum (RM-3439-PR)  (1962).


\end{thebibliography}
%

\providecommand{\noopsort}[1]{}\providecommand{\singleletter}[1]{#1}

\end{document}